\documentclass{sigchi}

\usepackage{balance}       
\usepackage{graphics}      
\usepackage[T1]{fontenc}   
\usepackage{txfonts}
\usepackage{mathptmx}
\usepackage[pdflang={en-US},pdftex]{hyperref}
\usepackage{color}
\usepackage{booktabs}
\usepackage{textcomp}
\usepackage{soul}
\usepackage{color, colortbl}
\usepackage{multirow}
\usepackage{cite}

\usepackage{microtype}        
\usepackage{ccicons}          

\usepackage{todonotes}

\def\plaintitle{ AVEID: Automatic Video System for Measuring Engagement In Dementia}

\def\emptyauthor{}
\def\plainkeywords{Engagement; Video-based analytics; Dementia}

\makeatletter
\def\url@leostyle{%
  \@ifundefined{selectfont}{
    \def\UrlFont{\sf}
  }{
    \def\UrlFont{\small\bf\ttfamily}
  }}
\makeatother
\def\pprw{8.5in}
\def\pprh{11in}

\definecolor{Black}{RGB}{0,0,0}
\definecolor{Gray}{gray}{0.9}

\newcolumntype{P}[1]{>{\centering\arraybackslash}p{#1}}
\newcolumntype{M}[1]{>{\centering\arraybackslash}m{#1}}

\definecolor{linkColor}{RGB}{6,125,233}
\hypersetup{%
  pdftitle={\plaintitle},
  pdfauthor={\emptyauthor},
  pdfkeywords={\plainkeywords},
  pdfdisplaydoctitle=true, 
  bookmarksnumbered,
  pdfstartview={FitH},
  colorlinks,
  citecolor=black,
  filecolor=black,
  linkcolor=black,
  urlcolor=linkColor,
  breaklinks=true,
  hypertexnames=false
}

\def\sharedaffiliation{%
\end{tabular}
\begin{tabular}{ccc}}
\makeatletter
\def\@copyrightspace{\relax}
\makeatother
\begin{document}
\title{\plaintitle}

\numberofauthors{3}
    \author{
Viral Parekh\thanks{indicates equal contribution}$^{*1}$, Pin Sym Foong\footnotemark[1]$^{*2}$, Shendong Zhao$^2$ and Ramanathan Subramanian$^3$
      \sharedaffiliation
      \affaddr{$^{1}$CVIT Lab} & \affaddr{$^{2}$NUS-HCI Lab} & \affaddr{$^{3}$University of Glasgow,}\\
      \affaddr{Int'l Institute of Information} & \affaddr{National University of Singapore,} &\affaddr{Singapore}\\
      \affaddr{Technology, Hyderabad} & \affaddr{Singapore} & \email{Ramanathan.Subramanian@glasgow.ac.uk}\\
      \email{viral@live.in} &\email{pinsym@u.nus.edu,} &\\
       &\email{zhaosd@comp.nus.edu.sg} &
          }


\teaser{ 
\begin{center}
\includegraphics[width=1\linewidth]{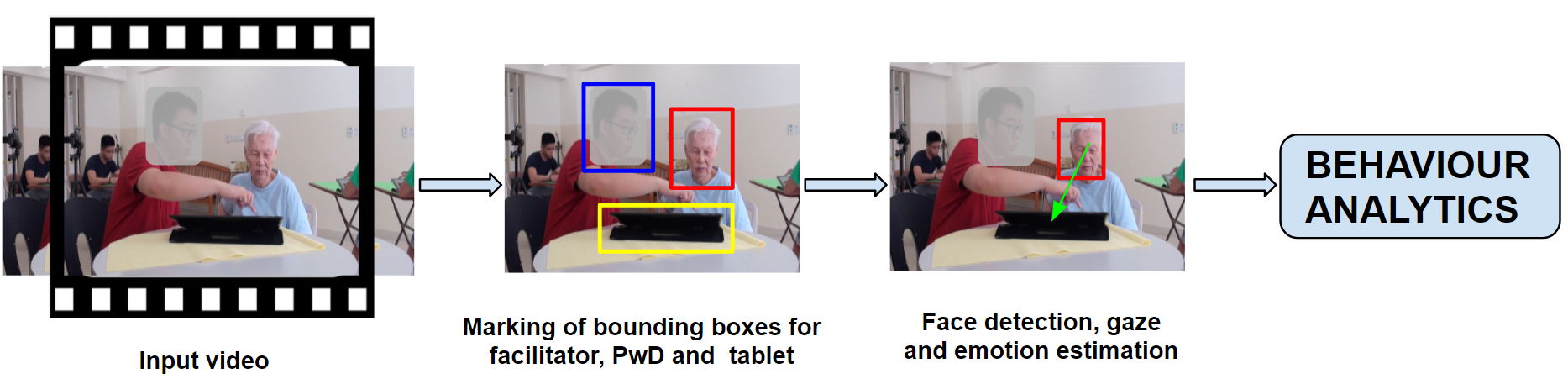}
\end{center}
\caption{AVEID overview - The AVEID system enables non-intrusive and automated behavioral analytics of persons with dementia (PwD) by capturing their attention (gazing behavior) and attitude (emotion) characteristics.}
\label{fig:avid_pipeline}

}

\maketitle

\begin{abstract}

Engagement in dementia is typically measured using behavior observational scales (BOS) that are tedious and involve intensive manual labor to annotate, and are therefore not easily scalable. We propose AVEID, a low cost and easy-to-use video-based engagement measurement tool to determine the engagement level of a person with dementia (PwD) during digital interaction. We show that the objective behavioral measures computed via AVEID correlate well with subjective expert impressions for the popular MPES and OME BOS, confirming its viability and effectiveness. Moreover, AVEID measures can be obtained for a variety of engagement designs, thereby facilitating large-scale studies with PwD populations.

\end{abstract}



\section{Introduction}
Engagement activities for people with dementia (PwD) are an important, non-pharmacological method to promote quality of life and reduce undesirable outcomes such as apathy, depression and aggressive behaviors \cite{r3}. Hence, HCI researchers have been developing various systems to supply more interactive and interesting engagement activities for PwD. Examples of these are conversation support systems \cite{r1}, art therapy \cite{r11} and music therapy \cite{r5}. Equally important are engagement measurement tools as they provide feedback to facilitators on the effectiveness of engagement systems, while also providing a basis for forming and adjusting interventions. As memory impairments in PwD preclude the use of self-reports as a measurement tool, researchers primarily use some form of observation to code outcome behavior. Behavioral coding requires the use of behavioral observational scales (BOS), and the training of coders who can accurately encode observed behaviors for robust inference.  Due to these requirements, behavioral coding as a measurement tool presents the following challenges: 1) It is human-effort intensive; 2) Training coders is time-consuming; 3) Large-scale data annotation becomes tedious, and 4) It supports only coarse-grained behavior analytics due to limitations in human annotation capability. 

Hence, automated measures of engagement might be useful for researchers. Despite the availability of such tools for neurotypical target groups, they are not appropriate for use with PwD. For example, PwD tend to resist any type of on-body physical instrumentation \cite{r10} so the use of wearable devices or bio-signal systems for measuring engagement is typically not viable. Patel et. al. \cite{r14} suggest that PwD require monitoring systems that are "unobtrusive, and preferably collected in a transparent way without patient intervention due to their cognitive impairment."

In this regard, we present AVEID, a low-cost and easy to use video-based system for measuring engagement in PwD. AVEID employs deep learning-based computer vision algorithms to continuously capture a dementia patient's engagement behavior during an interaction session, thereby enabling fine-grained behavior analytics. Consistent with BOS that quantify the patient's attention and attitude towards an engagement system, AVEID estimates the patient's attentional behavior based on gazing direction, and attitude based on facial emotions (Figs. \ref{fig:avid_pipeline},\ref{fig:avid_emotion}). {Gazing mannerisms have been extensively studied as cues indicative of attention/engagement during interactions~\cite{
Subramanian2010,Ricci_2015_ICCV}, while facial emotions are inherently reflective of a user's attitude towards the environment}. Also, since deep learning systems are `end-to-end' requiring no manual intervention for model synthesis, AVEID only requires manual annotation of bounding boxes to denote positions of the patient, facilitator (if present) and engagement device at the beginning of the examined video. Unlike gaze-tracking or wearable systems that involve specialized hardware, AVEID only requires a video recording as input. These features facilitate practical, day-to-day usage of AVEID in care homes by therapists or researchers from other domains.  

We validated AVEID against human (expert behavioral coder) impressions corresponding to two well-known BOS, namely, the Menorah Park Engagement Scale (MPES) \cite{r2} and the Observational Measure of Engagement (OME) \cite{r3}. Experiments confirm that measures derived from AVEID agree well with human opinion. Furthermore, AVEID can save the time and effort expended by the behavioral coder, allow for personalized treatment and enable timely analytics on large sample sizes. AVEID measures would also be applicable across small-space engagement activities, facilitating replicability and ecological validity of engagement evaluation studies; we ultimately envision AVEID to provide a strong basis for effective non-pharmacological intervention in dementia care environments.
\vspace{-0.5cm}
\section{AVEID Implementation}
Consistent with popular BOS used for measuring engagement with PwD, AVEID is designed to quantify the attention and attitude of the patient towards the engagement system over the observed period (Table 1). AVEID employs the patient's gaze focus as a cue towards inferring attention, while utilizing facial affect to infer attitude. We use the following terms to describe the system:
\begin{description}
	\setlength{\parskip}{0pt}
    \setlength{\itemsep}{0pt plus 1pt}
    \item [User:] who operates AVEID to measure engagement
    \item [Target subject:] PwD undertaking an engagement activity.
    \item [Target activity space:] the 2D area where we expect the PwD's gaze to be directed in order to engage with the designed activity. In the AVEID context, the engagement activity involves interaction with a tablet as in \cite{r7}. {The applications used in tablets are categorized by interests and abilities of the PwD. For interest areas, these five categories are selected reminiscence, household-linked activities, games, arts and crafts, and chatting.}
    \item [Facilitator:] a second person in the frame whose role is to support and promote engagement of the PwD.
\end{description}

\section{AVEID Modules}
The AVEID system comprises four modules, namely, \textit{User input, Face detection, Gaze and emotion detection and Behavior analytics} (Fig. \ref{fig:avid_pipeline}). 

The \textbf{User Input} module allows users to select the video for analysis, and enables them to mark bounding boxes corresponding to the target subject, target activity space and facilitator. This initialization needs to be performed at the beginning of each video for accurate face detection (under varied video acquisition conditions), and estimation of attentional measures. 

The \textbf{Face Detection} module implements the \textit{Tiny face} \cite{r8} state-of-the-art face detection method. Tiny face performs robust face detection across a wide range of illuminations, face sizes, head poses and facial occlusions. The face detection module detects patient's and facilitator's faces (within the input bounding boxes) in each video frame. 

\textbf{Gaze Detection} forms the core of AVEID, as its output is used to compute \textit{attention} measures, which are of prime importance in engagement measurement. This module implements the \textit{GazeFollow} deep network architecture of Recasens et al. \cite{r15}, and utilizes head orientation as a cue to determine where a target is gazing at \cite{r13,r16}.

\begin{figure}[!h]
\begin{center}
\includegraphics[width=0.9\linewidth]{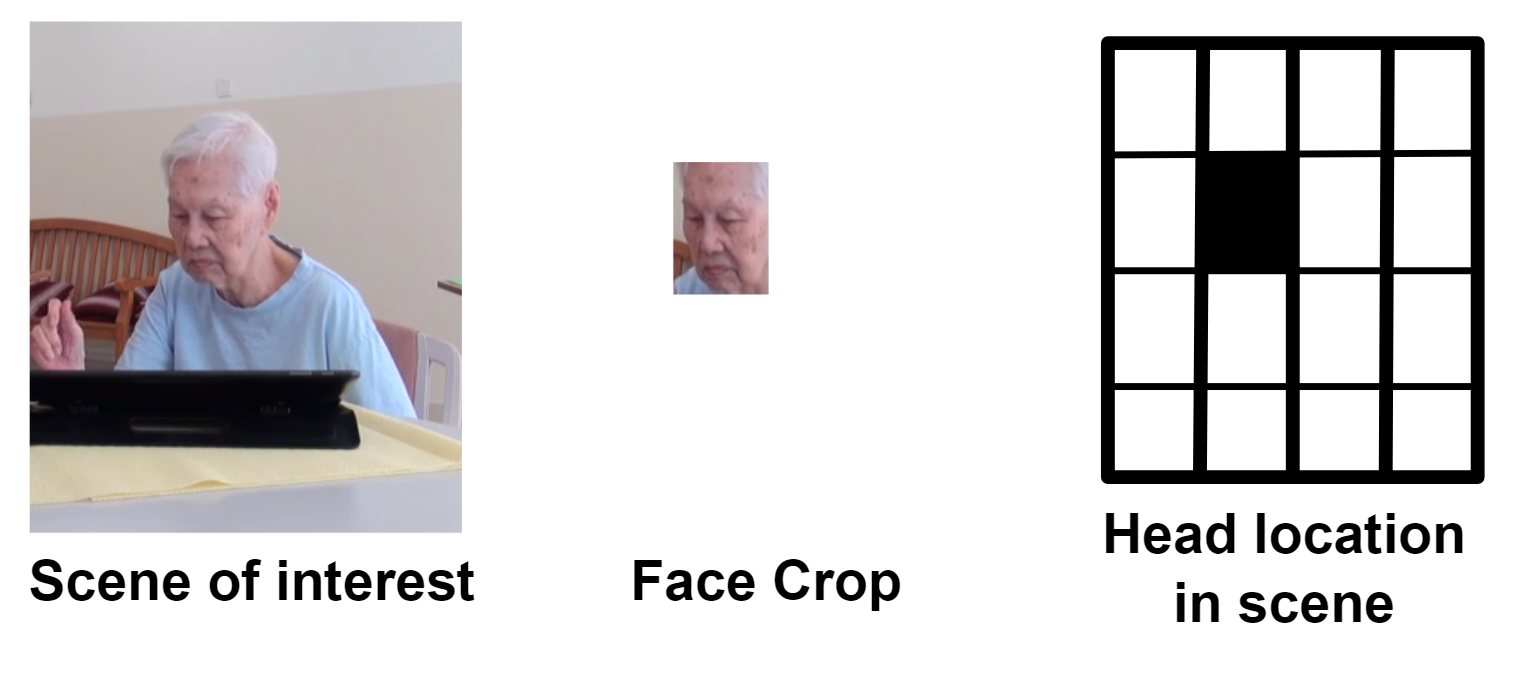}
\end{center}
\vspace{-.2cm}
\caption{The three inputs utilized for gaze detection.}
\label{fig:avid_input}
\vspace{-0.2cm}
\end{figure}

The target's gaze focus is determined based on three inputs (Fig. \ref{fig:avid_input}): 1) An image (video frame) capturing the scene of interest, 2) Cropped head of the target (output of \textit{Tiny face}), and 3) Location of the head in the scene (denoted by the highlighted grid square). The model comprises two computational pathways. The \textit{gaze pathway} uses the target head appearance and location to produce a \textit{gaze map} that estimates the general direction of the target's gaze. The \textit{saliency pathway} examines scene content to output the \textit{saliency map} that detects interesting objects capable of capturing the target's attention. The two maps are then combined to infer the target's gaze focus.  As the target activity space and facilitator are the two entities of interest in AVEID, the gaze detection module outputs for each video frame, a label signifying whether the target is gazing at the \textit{target activity space, facilitator or elsewhere}. 

The \textbf{Emotion Detection} module implements the deep network for emotion recognition described in \cite{r12}. Given that recognizing emotions of elderly people is challenging even for trained human experts, the deep network described in \cite{r12} is fine-tuned with 950 elderly face examples from the FACES dataset \cite{r4}. The emotion detection module outputs per video frame a label corresponding to one of the six Ekman emotions plus neutral as illustrated in Fig. \ref{fig:avid_emotion}.

\begin{table*}[!ht]
\fontsize{9}{9}\selectfont
\begin{center}
\renewcommand{\arraystretch}{1.3}
\begin{tabular}{M{20mm} M{45mm} M{50mm} M{45mm}}
\rowcolor{Black}
 & \textbf{\textcolor{white}{MPES BOS}} & \textbf{\textcolor{white}{OME BOS}} & \textbf{\textcolor{white}{AVEID}}\\
\rowcolor{Gray}
{\centering{\textbf{Unit of Assessment}}} & 
{\centering{5-minute observed periods, coded with 0,1 or 2}} & 
{\centering{Identified period of engagement, rated on a 7-point scale}} &
{\centering{User-specified observed periods of time (flexible granularity)}} \\
\rowcolor{Gray}
{\centering{\textbf{Attention}}} & 
{\centering{Active engagement (Did target activity), Passive engagement (Watched target activity), Other engagement.}} & 
{\centering{Attention intensity  (1 denoting no attention)}} & 
{\centering{3 raw + 18 derived gaze-based statistics over observed period.}}\\
\rowcolor{Gray}
{\centering{\textbf{Attitude}}} & 
{\centering{Pleasure and anxiety as proportion over observed period.}} & 
{\centering{Attitude valence with  (1 denoting strongly negative, 3 denoting neutral, and 7 denoting strongly positive affect).}} &
{\centering{Proportion of negative and neutral-or-positive affect over observed period.}} \\

\end{tabular}
\caption{Measuring attention and attitude via the MPES and OME scales, and the matching measures used with AVEID}
\label{table:avid_measures}
\end{center}
\vspace{-.3cm}
\end{table*}

The \textbf{Behavior Analytics} module processes the outputs of the gaze and emotion detection modules to compute measures reflecting the patient's \textit{attention} and attitude. For characterizing attention, the per-frame gaze labels are combined to compute three raw (or basic) statistics, namely, gaze proportion on tablet, facilitator and elsewhere over the period of observation. In addition to these coarse-grained features, we also derived 18 fine-grained statistics from the gaze labels for analysis, described as follows.

Upon determining \textit{episodes of focus} on the tablet, facilitator and other \textit{entities} within the observation period, we computed the means and standard deviations (std) of these episode durations (\textbf{6 features} in total); likelihood of transitioning from one entity to another- e.g., transition from focusing on tablet to focusing elsewhere; this gives rise to \textbf{six transition probability features} corresponding to 3 permute 2 entity transitions.  An additional \textit{six gaze flux features} denoting gaze flux into and out of the three entities were obtained from marginal likelihoods-- e.g., P(gaze flux into tab) = P(fac $\rightarrow$ tab) + P(others $\rightarrow$ tab), where $\rightarrow$ denotes a gaze transition.

To quantify attitude, we computed the proportions of positive (neutral or happy emotion) and negative (angry, sad or disgusted) affect over the observation period from the per-frame facial emotion labels. 

In terms of computational hardware, AVEID requires a Graphics Processing Unit (GPU) for video processing. The current system is implemented on a Xeon processor with 64GB RAM, and 12 GB NVIDIA GeForce GTX 1080 Ti GPU memory.

\section{BOS for Validation}
We validated AVEID by comparing the obtained attention and attitude measures against expert annotations acquired for the OME and MPES scales, whose descriptions follow. 

The \textbf{Observational Measure of Engagement} (OME) \cite{r3} is an observational scale to directly assess engagement in PwD. For this BOS, observers are first required to detect \textit{time periods denoting PwD engagement}, and score the attention and attitude levels of the patient within these engagement periods. OME represents a very coarse-grained assessment of PwD engagement, and can best facilitate examination of engagement periods, as no codings are made when the dementia patient is disengaged from the target activity. The \textbf{Menorah Park Engagement Scale} (MPES) \cite{r2} is a more fine-grained BOS, as PwD engagement is assessed over 5-minute time periods. Three types of engagement, namely, active engagement with the target device/activity, passive engagement and engagement with others, are measured in this BOS (Table \ref{table:avid_measures}).

\section{Expert Score acquisition}
All annotated videos were as shown in Fig.\ref{fig:avid_pipeline}, where a PwD engages with an interactive tablet aided by a facilitator \cite{r7}. For OME scoring, a dementia care therapist with 10 years experience indicated periods of patient engagement in seven 15-minute video segments according to the following OME definition: ``amount of attention the person was visibly paying to the stimulus (tablet) via eye movements; manipulating/holding and talking about it.'' \cite{r3}. MPES scores were provided by researchers trained to attain 0.8 (Kappa) inter-rater reliability. They scored 5-minute segments from 20 videos (30 minutes each), for active engagement (\textit{Did target activity}), passive engagement (\textit{Watched target activity}) and engagement with other stimuli (\textit{Attention on activity other than target}) on an ordinal scale (Table \ref{table:avid_measures}). Table \ref{table:avid_data} summarizes the annotation statistics.

\begin{table}[!h]
\fontsize{9}{9}\selectfont
\begin{center}
\renewcommand{\arraystretch}{1.3}
\begin{tabular}{c c c c}
\rowcolor{Black}
\textbf{\textcolor{white}{Bos}} & 
\multicolumn{1}{p{0.15\columnwidth}}{\centering{\textbf{\textcolor{white}{No. Videos}}}} &  
\multicolumn{1}{p{0.25\columnwidth}}{\centering{\textbf{\textcolor{white}{No. of Video Segments}}}} & 
\textbf{\textcolor{white}{Annotated by}}\\
\rowcolor{Gray}
\textbf{OME} & 7& \-- & Therapist \\
\rowcolor{Gray}
\textbf{MPES} & 20 & 130 & Trained researchers \\

\end{tabular}
\caption{Scores were obtained from experienced therapists or trained researchers.}
\label{table:avid_data}
\end{center}
\vspace{-.3cm}
\end{table}

\section{Validation Study Result and Discussion}\label{Res}
\subsection{Measures of Attention}
In the OME BOS, engagement periods are identified and then annotated for attention level and attitude. So, we computed the proportion of patient's gaze focus on the target activity space for a) those segments where the therapist indicated engagement, and b) the remaining video segments where the therapist inferred disengagement. Fig.3 presents the computed gaze proportions for seven videos. Higher distribution of gaze focus on tablet was clearly noted during engagement periods, as confirmed by a two-sample Kolmogorov-Smirnov test at p<0.001. Therefore, gaze focus on the target was sufficient to convey the notion of attention as with the OME.

The MPES BOS quantifies attention over 5-minute intervals. Fig. 4 presents Pearson correlations computed between active, passive and other engagement MPES scores, and the 21 AVEID attention features in the form of a 3$\times$21 grayscale image. Negative and positive correlations are respectively denoted by darker and lighter shades. Red and cyan symbols respectively denote significant (\textit{p}<0.05) and marginally significant (\textit{p}<0.1) correlations.

\begin{figure}[t]
\begin{center}
\includegraphics[width=1\linewidth]{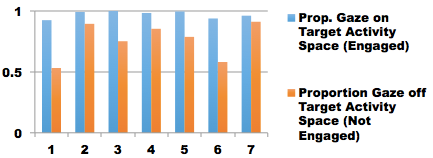}
\end{center}
\caption{Gaze proportions on tablet (engagement device) during \textit{engaged} and \textit{not-engaged} periods as per OME BOS.}
\label{fig:proportion}
\vspace{-0.3cm}
\end{figure}

Active engagement is significantly and positively correlated with the extent of gaze focused on tablet, mean duration of tablet gazing episodes and std of these episodic durations, while being negatively correlated with the extent of gaze focus on the caregiver and other areas. Active engagement is also marginally and positively correlated with the gaze flux in and out of the target activity area, suggesting that focus on target activity area as well as periodic gaze switching when communicating with the facilitator are linked with higher active engagement scores as assessed by the expert.

On the other hand, passive engagement marginally and negatively correlates with the gazing durations on the facilitator. This pattern is concurrent with the MPES description of passive activity where the PwD behaves with less enthusiasm and is not having social interactions with the facilitator. The final MPES item, engagement with other, positively correlates with gazing on facilitator, and negatively with the mean and standard deviation of gazing episode durations on the target activity area. This suggests that engagement with other activities, as coded by the experts, is associated with behaviors where the PwD directs attention more toward the facilitator rather than toward the presented activity. 

\begin{figure}[!hb]
\begin{center}
\includegraphics[width=1\linewidth]{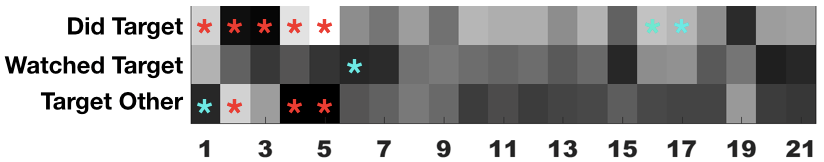}
\end{center}
\caption{Correlations between gaze-based AVEID features and MPES scores for active (Did Tgt), passive (Watched Tgt) and other engagement (Tgt Other). Red and cyan marks denote correlations significant at (\textit{p}<0.05) and (\textit{p}<0.1). }
\label{fig:correlation}
\vspace{-0.2cm}
\end{figure}

To summarize, gaze on target as a correlate of engagement is validated by both the OME and MPES coding methods. Furthermore, attention measures computed via AVEID are able to capture a number of aspects concerning these BOS.

\subsection{Measures of Attitude}
Inferring facial emotions of PwD is a known challenge due to older adults exhibiting facial emotions in a controlled manner, ageing skin and muscles significantly modulating facial appearance, and indicating a prominently negative affect \cite{r6}. Additionally, PwD often display flattened affect \cite{r9}. However, greater engagement from the PwD should also elicit a positive reaction from the facilitator whose role is to promote such behavior. As facilitators are neurotypical adults whose facial emotions can be better recognized with available computer vision tools, we therefore examined if higher MPES attitude scores correlate better with the facilitator's facial emotions (Fig. \ref{fig:avid_emotion}). This hypothesis turned out to be true, with pleasure scores correlating significantly and positively with the proportion of positive facial affect exhibited by the facilitator (\textit{r}=0.24, \textit{p}<0.01).{Therefore, examining facilitator behavior could provide crucial cues for measuring engagement in PwD.} 


Overall, results reveal that AVEID can effectively capture patient (and facilitator) behavior indicative of attention and attitude. Also, since AVEID measures are based on per-frame gaze and emotion labels, it is possible to go beyond coarse engagement measures that BOS provide. E.g., even though the patient's verbal behavior was not captured in the videos, the frequency with which the patient directs gaze towards the facilitator may serve as an effective cue to this end. Finally, gazing and attitude estimation can be reliably accomplished for small-space activities (where the patient's face is clearly visible), facilitating evaluation of multiple engagement designs.   

\section{Challenges and Limitations}
\begin{figure}[hb]
\begin{center}
\includegraphics[width=1\linewidth]{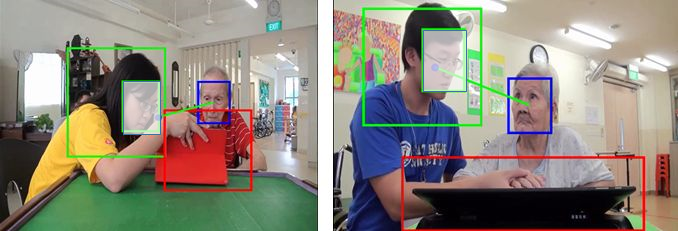}
\end{center}
\caption{Incorrect gaze estimation examples (zoom to view).}
\label{fig:avid_gaze}
\vspace{0.1cm}
\begin{center}
\includegraphics[width=1\linewidth]{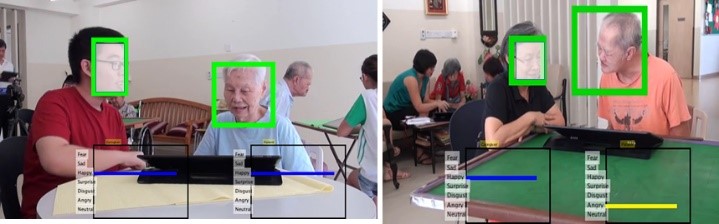}
\end{center}
\caption{Exemplar emotion estimation results. Facial emotions of the PwD and facilitator are correctly identified (left). PwD's emotion is incorrect, but facilitator's emotion is correct (right) (zoom to view).}
\vspace{-.3cm}
\label{fig:avid_emotion}
\end{figure}

Figures \ref{fig:avid_gaze} and \ref{fig:avid_emotion} illustrate the challenges involved in video-based engagement measurement for PwD, where the observation videos are captured under unconstrained settings. Fig.\ref{fig:avid_gaze} presents two examples of incorrect gaze focus estimation due to the closeness of the facilitator to the tablet, and due to the 2D video information being insufficient to model the 3D world. Likewise, in Fig. \ref{fig:avid_emotion} (right), the patient's facial appearance is mistaken by the algorithm as an exhibition of sadness.{It needs to be acknowledged that gazing behavior can only indicate passive engagement; the use of wearable physiological sensors~\cite{ASCERTAIN} may be necessary for inferring cognitive engagement.} 

\section{Conclusion}
We present AVEID, a video-based analytics system that effectively capture various aspects of BOS employed for measuring engagement in PwD.{AVEID currently measures only passive engagement (via gaze behavior), and inferring attitude from PwDs' facial emotions is highly challenging.} Future work will address these limitations, and explore additional modalities (such as verbal behavior) for measuring engagement among PwD.

\balance{}

\bibliographystyle{SIGCHI-Reference-Format}
\bibliography{aveid}


\begin{thebibliography}{00}


\ifx \showCODEN    \undefined \def \showCODEN     #1{\unskip}     \fi
\ifx \showDOI      \undefined \def \showDOI       #1{{\tt DOI:}\penalty0{#1}\ }
  \fi
\ifx \showISBNx    \undefined \def \showISBNx     #1{\unskip}     \fi
\ifx \showISBNxiii \undefined \def \showISBNxiii  #1{\unskip}     \fi
\ifx \showISSN     \undefined \def \showISSN      #1{\unskip}     \fi
\ifx \showLCCN     \undefined \def \showLCCN      #1{\unskip}     \fi
\ifx \shownote     \undefined \def \shownote      #1{#1}          \fi
\ifx \showarticletitle \undefined \def \showarticletitle #1{#1}   \fi
\ifx \showURL      \undefined \def \showURL       #1{#1}          \fi

\bibitem{r1}
{Arlene~J Astell}, {Maggie~P Ellis}, {Lauren Bernardi}, {Norman Alm}, {Richard
  Dye}, {Gary Gowans}, {and} {Jim Campbell}. 2010.
\newblock \showarticletitle{Using a touch screen computer to support
  relationships between people with dementia and caregivers}.
\newblock {\em Interacting with Computers\/} {22}, 4 (2010), 267--275.
\newblock


\bibitem{r3}
{Jiska Cohen-Mansfield}, {Maha Dakheel-Ali}, {and} {Marcia~S Marx}. 2009.
\newblock \showarticletitle{Engagement in persons with dementia: the concept
  and its measurement}.
\newblock {\em The American journal of geriatric psychiatry\/} {17}, 4 (2009),
  299--307.
\newblock


\bibitem{r4}
{Natalie~C Ebner}, {Michaela Riediger}, {and} {Ulman Lindenberger}. 2010.
\newblock \showarticletitle{FACES-A database of facial expressions in young,
  middle-aged, and older women and men: Development and validation}.
\newblock {\em Behavior research methods\/} {42}, 1 (2010), 351--362.
\newblock


\bibitem{r5}
{Stu Favilla} {and} {Sonja Pedell}. 2013.
\newblock \showarticletitle{Touch screen ensemble music: collaborative
  interaction for older people with dementia}. In {\em Proceedings of the 25th
  Australian Computer-Human Interaction Conference: Augmentation, Application,
  Innovation, Collaboration}. ACM, 481--484.
\newblock


\bibitem{r6}
{Mara F\"olster}, {Ursula Hess}, {and} {Katja Werheid}. 2014.
\newblock \showarticletitle{Facial age affects emotional expression decoding}.
\newblock {\em Frontiers in Psychology\/}  {5} (2014), 30.
\newblock
\showISSN{1664-1078}
\showDOI{%
\url{http://dx.doi.org/10.3389/fpsyg.2014.00030}}


\bibitem{r7}
{Pin~Sym Foong}, {Shengdong Zhao}, {Kelsey Carlson}, {and} {Zhe Liu}. 2017.
\newblock \showarticletitle{VITA: Towards Supporting Volunteer Interactions
  with Long-Term Care Residents with Dementia}. In {\em Proceedings of the 2017
  CHI Conference on Human Factors in Computing Systems} {\em (CHI '17)}. ACM,
  New York, NY, USA, 6195--6207.
\newblock
\showISBNx{978-1-4503-4655-9}
\showDOI{%
\url{http://dx.doi.org/10.1145/3025453.3025776}}


\bibitem{r8}
{Peiyun Hu} {and} {Deva Ramanan}. 2017.
\newblock \showarticletitle{Finding tiny faces}. In {\em 2017 IEEE Conference
  on Computer Vision and Pattern Recognition (CVPR)}. IEEE, 1522--1530.
\newblock


\bibitem{r2}
{Cameron J.Camp}, {Michael J.~Skrajner}, {and} {Greg~J. Gorzelle}.
\newblock {\em Engagement in Dementia. In Assessment Scales for Advanced
  Dementia}.
\newblock HPP, Baltimore. 65--69 pages.
\newblock


\bibitem{r9}
{M~Powell Lawton}, {Kimberly Van~Haitsma}, {Margaret Perkinson}, {and} {Katy
  Ruckdeschel}. 1999.
\newblock \showarticletitle{Observed affect and quality of life in dementia:
  Further affirmations and problems.}
\newblock {\em Journal of mental Health and Aging\/} (1999).
\newblock


\bibitem{r10}
{Edward~L Mahoney} {and} {Diane~F Mahoney}. 2010.
\newblock \showarticletitle{Acceptance of wearable technology by people with
  Alzheimer’s disease: Issues and accommodations}.
\newblock {\em American Journal of Alzheimer's Disease \& Other
  Dementias{\textregistered}\/} {25}, 6 (2010), 527--531.
\newblock


\bibitem{r11}
{Alex Mihailidis}, {Scott Blunsden}, {Jennifer Boger}, {Brandi Richards},
  {Krists Zutis}, {Laurel Young}, {and} {Jesse Hoey}. 2010.
\newblock \showarticletitle{Towards the development of a technology for art
  therapy and dementia: Definition of needs and design constraints}.
\newblock {\em The Arts in Psychotherapy\/} {37}, 4 (2010), 293--300.
\newblock


\bibitem{r12}
{Hong-Wei Ng}, {Viet~Dung Nguyen}, {Vassilios Vonikakis}, {and} {Stefan
  Winkler}. 2015.
\newblock \showarticletitle{Deep Learning for Emotion Recognition on Small
  Datasets Using Transfer Learning}. In {\em Proceedings of the 2015 ACM on
  International Conference on Multimodal Interaction} {\em (ICMI '15)}. ACM,
  New York, NY, USA, 443--449.
\newblock
\showISBNx{978-1-4503-3912-4}
\showDOI{%
\url{http://dx.doi.org/10.1145/2818346.2830593}}


\bibitem{r13}
{Viral Parekh}, {Ramanathan Subramanian}, {and} {C.~V. Jawahar}. 2017.
\newblock {\em Eye Contact Detection via Deep Neural Networks}.
\newblock Springer International Publishing, Cham, 366--374.
\newblock
\showISBNx{978-3-319-58750-9}
\showDOI{%
\url{http://dx.doi.org/10.1007/978-3-319-58750-9_51}}


\bibitem{r14}
{Shyamal Patel}, {Hyung Park}, {Paolo Bonato}, {Leighton Chan}, {and} {Mary
  Rodgers}. 2012.
\newblock \showarticletitle{A review of wearable sensors and systems with
  application in rehabilitation}.
\newblock {\em Journal of NeuroEngineering and Rehabilitation\/} {9}, 1 (20 Apr
  2012), 21.
\newblock
\showISSN{1743-0003}
\showDOI{%
\url{http://dx.doi.org/10.1186/1743-0003-9-21}}


\bibitem{r15}
{Adria Recasens$^*$}, {Aditya Khosla$^*$}, {Carl Vondrick}, {and} {Antonio
  Torralba}. 2015.
\newblock \showarticletitle{Where are they looking?}. In {\em Advances in
  Neural Information Processing Systems (NIPS)}.
\newblock
\newblock
\shownote{$^*$ indicates equal contribution.}


\bibitem{Ricci_2015_ICCV}
{Elisa Ricci}, {Jagannadan Varadarajan}, {Ramanathan Subramanian}, {Samuel
  Rota~Bulo}, {Narendra Ahuja}, {and} {Oswald Lanz}. 2015.
\newblock \showarticletitle{Uncovering Interactions and Interactors: Joint
  Estimation of Head, Body Orientation and F-Formations From Surveillance
  Videos}. In {\em {ICCV}}.
\newblock


\bibitem{r16}
{Brian~A. Smith}, {Qi Yin}, {Steven~K. Feiner}, {and} {Shree~K. Nayar}. 2013.
\newblock \showarticletitle{Gaze Locking: Passive Eye Contact Detection for
  Human-object Interaction}. In {\em Proceedings of the 26th Annual ACM
  Symposium on User Interface Software and Technology} {\em (UIST '13)}. ACM,
  New York, NY, USA, 271--280.
\newblock
\showISBNx{978-1-4503-2268-3}
\showDOI{%
\url{http://dx.doi.org/10.1145/2501988.2501994}}


\bibitem{Subramanian2010}
{Ramanathan Subramanian}, {Jacopo Staiano}, {Kyriaki Kalimeri}, {Nicu Sebe},
  {and} {Fabio Pianesi}. 2010.
\newblock \showarticletitle{Putting the Pieces Together: Multimodal Analysis of
  Social Attention in Meetings}. In {\em {ACM} Multimedia}. 659--662.
\newblock


\bibitem{ASCERTAIN}
{R. Subramanian}, {J. Wache}, {M. Abadi}, {R. Vieriu}, {S. Winkler}, {and} {N.
  Sebe}. 2017.
\newblock \showarticletitle{ASCERTAIN: Emotion and Personality Recognition
  using Commercial Sensors}.
\newblock {\em IEEE Transactions on Affective Computing\/} (2017).
\newblock


\end{thebibliography}

\end{document}